\documentclass[prd,aps,10pt,nofootinbib,twocolumn,superscriptaddress,preprintnumbers,balancelastpage,longbibliography]{revtex4-1}

\usepackage{dcolumn}
\usepackage[table]{xcolor}
\usepackage{youngtab}
\usepackage{ytableau}

\usepackage{booktabs}    
\usepackage{placeins}
\usepackage{soul}

\definecolor{redd}{rgb}{0.8, 0.1,0.2}
\definecolor{navy}{rgb}{0.05, 0.23,0.75}
\usepackage[colorlinks=true,citecolor=red,linkcolor=blue]{hyperref}
\usepackage{chngcntr}
\hypersetup{
     colorlinks   = true,
     citecolor    = navy,
	linkcolor = redd,
	urlcolor=navy,
	anchorcolor=blue
}
\usepackage{bbm}
\usepackage{amsfonts}
\usepackage{amsmath,amssymb}
\usepackage{mathrsfs}
\usepackage{epsfig}
\usepackage{graphicx}
\usepackage{wrapfig}                 
\usepackage{url}
\usepackage{hyperref}
\usepackage{float}
\usepackage{color}
\usepackage{multirow}
\usepackage{lipsum}
\usepackage{enumitem}
\usepackage{ctable}
\newcolumntype{L}{>{\centering\arraybackslash}m{1.5cm}}

\usepackage{enumitem}
\usepackage{chngcntr}

\usepackage{makecell}

\newcommand{\be}{\begin{equation}}
\newcommand{\ee}{\end{equation}}
\newcommand{\bea}{\begin{eqnarray}}
\newcommand{\eea}{\end{eqnarray}}
\newcommand{\bc}{\begin{center}}
\newcommand{\ec}{\end{center}}

\begin{document}
		
\title{
High-Temperature Vacua of the Standard Model from One-Form Symmetry
}

\author{Ling-Xiao Xu}
\email{phy.lingxiao.xu@gmail.com}
\affiliation{Abdus Salam International Centre for Theoretical Physics, Strada Costiera 11, 34151, Trieste, Italy}

\begin{abstract}
We study the high-temperature vacua of the Standard Model from the perspective of generalized symmetries and the global structure of the gauge group. We analyze the one-loop effective potential for Polyakov loop holonomies, incorporating the gauge identifications induced by gauging discrete subgroups of the electric one-form symmetry. This provides a classification of physically inequivalent vacua for each gauge-group quotient. We further comment on the effects of additional center-charged matter on the residual one-form symmetry and the vacuum structure.
\end{abstract}

\maketitle

\section{Introduction}
\label{sec:intro}

The global structure of a gauge theory carries physical information beyond its local particle content and interactions~\cite{Aharony:2013hda, Aharony:2013kma}. In the Standard Model (SM), the gauge algebra $\mathfrak{su}(3)_c\oplus\mathfrak{su}(2)_L\oplus\mathfrak{u}(1)_Y$ does not uniquely determine the connected gauge group. The possible global forms are
\begin{equation}
    G_\Gamma=\frac{SU(3)_c\times SU(2)_L\times U(1)_Y}{\Gamma}, \qquad \Gamma\subseteq\mathbb{Z}_6,
    \label{eq:SM_global}
\end{equation}
all of which are compatible with the quantum numbers of the SM particles. These choices differ in their allowed
line operators~\cite{Tong:2017oea}, providing observables sensitive to the global gauge structure beyond local particle excitations. (For an overview of recent developments from complementary perspectives and further references, see~\cite{Davighi:2019rcd, Davighi:2020bvi, Wan:2019gqr, Wang:2021ayd, Wang:2020mra, Anber:2021upc, Choi:2023pdp, Reece:2023iqn, Cordova:2023her, vanBeest:2023mbs, Li:2024nuo, Alonso:2024pmq, Koren:2024xof, Dierigl:2024cxm, Cao:2024lwg, Khoze:2024hlb, Alonso:2025rkk, Anber:2025gvb, Herrero-Brocal:2025rkx, Delgado:2026lwj, Vecchi:2025qie, Hamada:2025cwu, Hsin:2024lya, Wan:2024kaf}.)

Compactification makes this distinction particularly transparent~\cite{Aharony:2013hda, Aharony:2013kma}. Line operators wrapping the compact direction become local operators in the lower-dimensional theory, whose expectation values can distinguish phases and expose differences in vacuum structure (see~\cite{Poppitz:2021cxe, Fukushima:2017csk} and the references therein). For a thermal circle, wrapped Wilson lines are Polyakov loops, and the electric one-form symmetry~\cite{Gaiotto:2014kfa} gives rise to a zero-form center symmetry acting on them. The component acting on spatial Wilson lines remains a one-form symmetry.
For the SM product group $G_{\{1\}}$, the electric one-form symmetry is $\mathbb{Z}_6$. Gauging a subgroup $\Gamma$ restricts the genuine Polyakov loop observables and identifies holonomies related by the corresponding center transformations. Therefore, the global gauge structure determines the physical \emph{holonomy domain} on which the SM thermal vacuum structure should be analyzed.

The one-loop thermal holonomy potential~\cite{Gross:1980br, Weiss:1980rj, Weiss:1981ev} of the SM and its global and metastable minima have been studied previously~\cite{KorthalsAltes:1994be, Bronoff:1997an}. 
However, the physical configuration space of the holonomies depends on the global form of the gauge group, an aspect that has not been incorporated in these analyses. In this work, we construct the physical holonomy domains for all four global forms $G_\Gamma$ of the SM gauge group by explicitly implementing the gauge identifications associated with gauging the $\Gamma\subset \mathbb{Z}_6$ of the electric one-form symmetry. While these different global forms leave the local thermal potential unchanged, they change the physical holonomy space and hence the classification of the gauge-inequivalent high-temperature vacua. In particular, the six center-related minima of the product group are reduced to $6/|\Gamma|$ physically distinct vacua. Furthermore, we briefly examine how additional heavy particles modify the vacuum structure.

\section{Global Gauge Structure and the Physical Holonomy Domain}
\label{sec:holonomy}

\noindent\textbf{The trivial quotient.---} We first consider the product group $G_{\{1\}}=SU(3)_c\times SU(2)_L\times U(1)_Y$. For constant Cartan backgrounds, the thermal holonomy is
\begin{equation}
\Omega(q,r,s,t)=\exp\!\left\{\frac{i}{T}
\left(g_{st}\mathcal A_0+gA_0+\frac{g'}6 B_0\right)\right\}.
\label{eq:thermal_holo}
\end{equation}
where the thermal circle has length $1/T$, and $\mathcal{A}_0$, $A_0$, $B_0$ denote the temporal color, weak, and hypercharge gauge fields with gauge couplings $g_{st}, g, g^\prime$, respectively. The hypercharge is normalized to $Y=1/6$. We parameterize the background as~\cite{KorthalsAltes:1994be}
\begin{equation}
\mathcal A_0=\frac{2\pi T}{g_{st}}
\begin{pmatrix}
q/3+r/2&0&0\\0&q/3-r/2&0\\0&0&-2q/3
\end{pmatrix}
\end{equation}
\begin{equation}
A_0=\frac{2\pi T}{g}
\begin{pmatrix}s/2&0\\0&-s/2\end{pmatrix},\qquad
B_0=\frac{2\pi T}{g'}\,t.
\label{eq:cartan-backgrounds}
\end{equation}
For the product group $G_{\{1\}}$, the gauge identifications act independently on the color, weak, and hypercharge holonomies. Following~\cite{Anber:2015wha}, we first determine the fundamental domain under large gauge transformations and then impose Weyl equivalence.

\begin{figure*}[t]
    \centering
    \includegraphics[width=\textwidth]
    {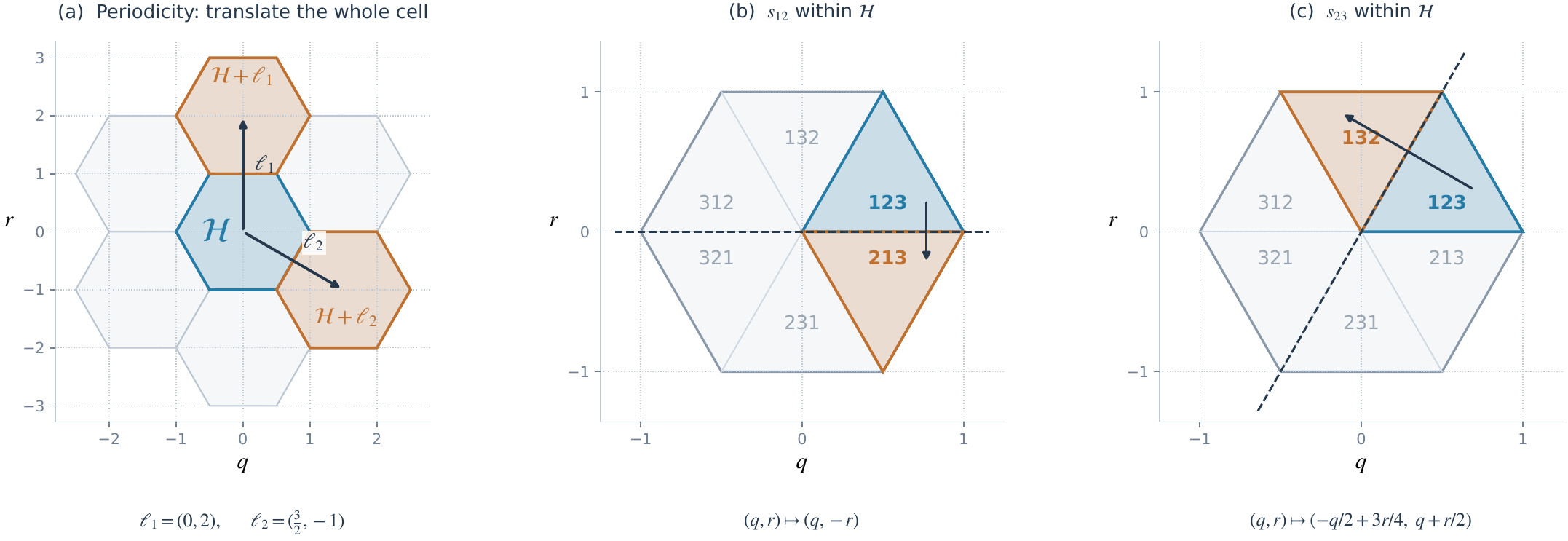}
    \caption{Co-root lattice and Weyl identifications of the $SU(3)_c$ thermal holonomy.
    (a) The translations $\ell_1=(0,2)$ and $\ell_2=(3/2,-1)$ relate hexagonal fundamental cells.
    (b,c) The Weyl generators $s_{12}$ and $s_{23}$ preserve the hexagon and identify its six ordering sectors, leaving one triangle domain (i.e., the Weyl chamber), where the indices label the eigenvalues of $\frac{g_{st}}{2\pi T} \mathcal A_0$ in descending order. For example, the Weyl chamber denoted by $123$ corresponds to the ordering $q/3+r/2\geq q/3-r/2\geq-2q/3$. (Notice that the Weyl chamber will be further reduced to a smaller regime for $SU(3)_c/\mathbb{Z}_3$. However, since the SM quarks are in the fundamental representation of $SU(3)_c$, this is not a viable option for us.)
    }
    \label{fig:color-holonomy-domain}
\end{figure*}

For $SU(3)_c$, large gauge transformations periodic in the group shift the three eigenphases by integers whose
sum vanishes. We therefore choose independent shifts for the three eigenvalues of $\frac{g_{st}}{2\pi T} \mathcal A_0$ as $(1,-1,0)$ and $(0,1,-1)$, which lead to the identifications
\begin{equation}
    (q,r)\sim(q,r+2),\qquad
    (q,r)\sim\left(q+\frac32,r-1\right).
\end{equation}
A fundamental cell of the co-root lattice is chosen to be the hexagonal Voronoi cell $\mathcal{H}$ with opposite edges identified by lattice translations, as illustrated in Fig.~\ref{fig:color-holonomy-domain}(a).
Furthermore, the Weyl group $S_3$ permutes the three color eigenphases, holonomies related by these permutations are gauge equivalent and must therefore also be identified. The generators of $S_3$ can be chosen to exchange the first two and the last two eigenvalues, acting as
\begin{equation}
\begin{split}
    s_{12}:&\quad(q,r)\mapsto(q,-r),\\
    s_{23}:&\quad(q,r)\mapsto
    \left(-\frac q2+\frac{3r}{4},\,q+\frac r2\right). \label{eq:weyl_s23}
\end{split}
\end{equation}
Both preserve $\mathcal H$ and together identify its six ordering sectors; see Fig.~\ref{fig:color-holonomy-domain}(b,c). Choosing $q/3+r/2\geq q/3-r/2\geq-2q/3$ imposes $r\geq0$ and $q\geq r/2$. Within this sector, the hexagonal bounds reduce to $q+r/2\leq1$, yielding the Weyl chamber,
\begin{equation}
    0\leq r\leq1,\qquad
    \frac r2\leq q\leq1-\frac r2.
    \label{eq:color-chamber}
\end{equation}

For $SU(2)_L$, the eigenvalues of $\frac{g}{2\pi T} A_0$ are $\pm s/2$. Large gauge transformations periodic in $SU(2)_L$ identify $s\sim s+2$, giving the initial domain $0\leq s<2$. The Weyl group $\mathbb{Z}_2$ exchanges the eigenvalues,
$s\mapsto-s$, further identifying $s$ with $2-s$ within $0\leq s<2$. Consequently, the Weyl chamber for $SU(2)_L$ is~\footnote{For comparison, for $SO(3)\simeq SU(2)/\mathbb{Z}_2$, there is the additional identification $s\sim s+1$ because the $\mathbb{Z}_2$ center acts trivially, which yields $s\sim 1-s$ within the $SU(2)$ Weyl chamber combining with $s\sim -s$. Consequently, the Weyl chamber of $SU(2)/\mathbb{Z}_2$ is reduced to $0\leq s\leq\tfrac{1}{2}$~\cite{Smilga:1993vb, Smilga:1996cm, Cheluvaraja:1998yu, Poppitz:2021cxe}.}
\begin{equation}
    0\leq s\leq1.
    \label{eq:weak-chamber}
\end{equation}
Finally, the hypercharge holonomy is $e^{i\pi t/3}$. Its periodicity requires $t\sim t+6$, giving
\begin{equation}
0\leq t<6.
\label{eq:hypercharge-domain}
\end{equation}
There is no Weyl identification for the Abelian factor.

\begin{figure*}[t]
    \centering
    \includegraphics[width=\textwidth]{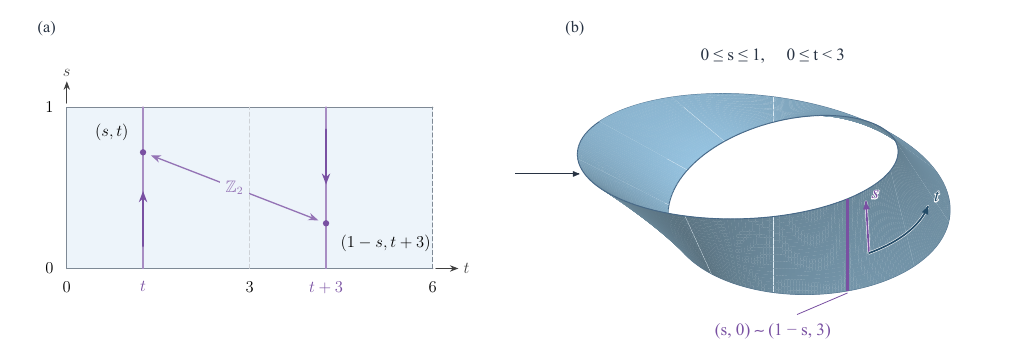}
    \caption{Weak--hypercharge holonomy identification for $G_{\mathbb Z_2}$. (a) Within the product-group domain $0\le s\le1$, $0\le t<6$, the quotient pairs $(s,t)$ with $(1-s,t+3)$ for $0\le t<3$; opposite arrows indicate the reversal of the weak coordinates.
    (b) The quotient space is a M\"obius band, represented by $0\le s\le1$, $0\le t<3$, where the $\mathbb{Z}_2$ quotient identifies the points $(s,0)\sim (1-s, 3)$ at the boundary of $t$ being glued together. The arrows indicate the local directions of increasing $s$ and $t$.}
    \label{fig:sm-z2-domain}
\end{figure*}

\noindent\textbf{The $\mathbb{Z}_2$ quotient.---} For $G_{\mathbb{Z}_2}$, the quotient is generated by $(\mathbf{1}_3,-\mathbf{1}_2,-1)$ and induces the joint identification between the weak and hypercharge holonomies
\begin{equation}
    \left(e^{\pm i\pi s}, e^{i\pi t/3}\right) \sim \left(-e^{\pm i\pi s}, -e^{i\pi t/3}\right)\;.
\end{equation}
In the Cartan coordinates, this identification is 
\begin{equation}
    (q,r,s,t)\sim(q,r,s+1,t+3).
    \label{eq:z2_identification_gauge}
\end{equation}
The weak Weyl reflection and periodicity return $1+s$ to $1-s$ within the weak Weyl chamber, i.e., $s+1\sim-(s+1)+2=1-s$. The quotient therefore pairs the two halves of the product-group domain as~\footnote{For $3\leq t<6$, the periodicity $t\sim t+6$ gives $(q,r,s,t)\sim(q,r,1-s,t+3)\sim (q, r,1-s,t-3)$. Since $0\leq t-3<3$ and $1-(1-s)=s$, we recognize that this is the same identification condition in Eq.~\eqref{eq:sm-z2-identification}.}
\begin{equation}
    (q,r,s,t)\sim(q,r,1-s,t+3),
    \qquad 0\leq t<3.
    \label{eq:sm-z2-identification}
\end{equation}
Therefore, we may retain the full color and weak Weyl chambers and restrict the hypercharge coordinate to obtain
\begin{equation}
    0\leq r\leq1,\quad
    \frac r2\leq q\leq1-\frac r2,\quad
    0\leq s\leq1,\quad
    0\leq t<3.
    \label{eq:sm-z2-domain}
\end{equation}
The boundaries at $t=0$ and $t=3$ are glued with a twist for $s$, i.e., $(q,r,s,0)\sim(q,r,1-s,3)$, making the weak-hypercharge holonomy space a M\"obius band while the color Weyl chamber is unchanged, as illustrated in Fig.~\ref{fig:sm-z2-domain}.

\begin{figure*}[t]
    \centering
    \includegraphics[width=\textwidth]{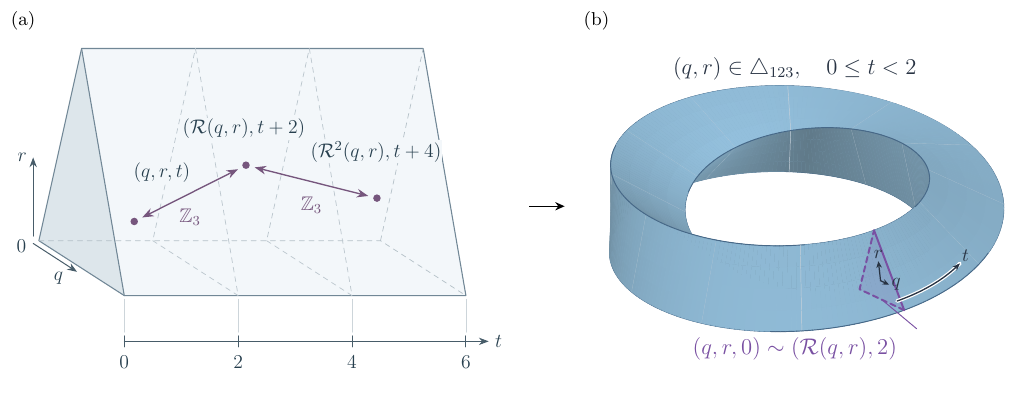}
    \caption{Color--hypercharge holonomy domain for $G_{\mathbb Z_3}$. (a) The product-group domain, where the three marked representatives are identified by the $\mathbb{Z}_3$ quotient, where the map $\mathcal R$ on the $(q,r)$ plane is accompanied by $t\mapsto t+2$. 
    (b) Retaining $0\leq t<2$ and gluing the boundary faces by $\mathcal R$ gives a solid torus, here the $(q,r)$ are within the color Weyl chamber. (This is analogous to the M\"obius band in the $G_{\mathbb Z_2}$ case.)}
    \label{fig:sm-z3-domain}
\end{figure*}

\noindent\textbf{The $\mathbb Z_3$ quotient.---} For $G_{\mathbb Z_3}$, the quotient is generated by $(\omega^2\mathbf{1}_3,\mathbf{1}_2,\omega)$, where $\omega=e^{2\pi i/3}$, and induces $(q,r,s,t)\sim(q+2,r,s,t+2)$. Using the color periodicity $q\sim q+3$, we replace $q+2$ by $q-1$, which translates the $123$ triangle (i.e., the color Weyl chamber) into the $312$ triangle within the hexagonal cell; see in Fig.~\ref{fig:color-holonomy-domain}. The Weyl transformations $s_{23}$ and $s_{12}$ then maps the $312$ triangle back to the original $123$ triangle. Namely, from Eq.~\eqref{eq:weyl_s23}, we obtain $\mathcal R(q,r)=s_{12}(s_{23}(q,r))$, i.e.,
\begin{equation}
    \mathcal R(q,r)= 
    \left(\frac{1}{2}-\frac{q}{2}+\frac{3r}{4},\,
    1-q-\frac r2\right),
    \quad \mathcal R^3=1.
    \label{eq:sm-z3-color-map}
\end{equation}
The quotient therefore identifies three representatives in the product-group domain,~\footnote{For
$2j\leq t<2(j+1)$, with $j=1,2$, the inverse quotient transformation gives $(q,r,s,t)\sim(\mathcal R^{-j}(q,r),s,t-2j)$. Since $0\leq t-2j<2$ and $\mathcal R^3=1$, these are the same identifications as in Eq.~\eqref{eq:sm-z3-identification}.}
\begin{equation}
\begin{split}
    (q,r,s,t)
    &\sim(\mathcal R(q,r),s,t+2)\\
    &\sim(\mathcal R^2(q,r),s,t+4),
    \qquad 0\leq t<2,
\end{split}
\label{eq:sm-z3-identification}
\end{equation}
where $R^2(q,r)$ denotes the inverse map of $R(q,r)$.
Consequently, we retain the full color and weak Weyl chambers while restricting the hypercharge coordinate:
\begin{equation}
    0\leq r\leq1,\quad \frac r2\leq q\leq1-\frac r2, \quad 0\leq s\leq1,\quad 0\leq t<2.
\label{eq:sm-z3-domain}
\end{equation}
The boundary surfaces are glued as $(q,r,s,0)\sim(\mathcal R(q,r),s,2)$.
This cyclic gluing makes the color--hypercharge holonomy space a solid torus, as illustrated in Fig.~\ref{fig:sm-z3-domain}, while the weak chamber remains unchanged.

\begin{figure*}[t]
    \centering
    \includegraphics[width=0.8\textwidth]{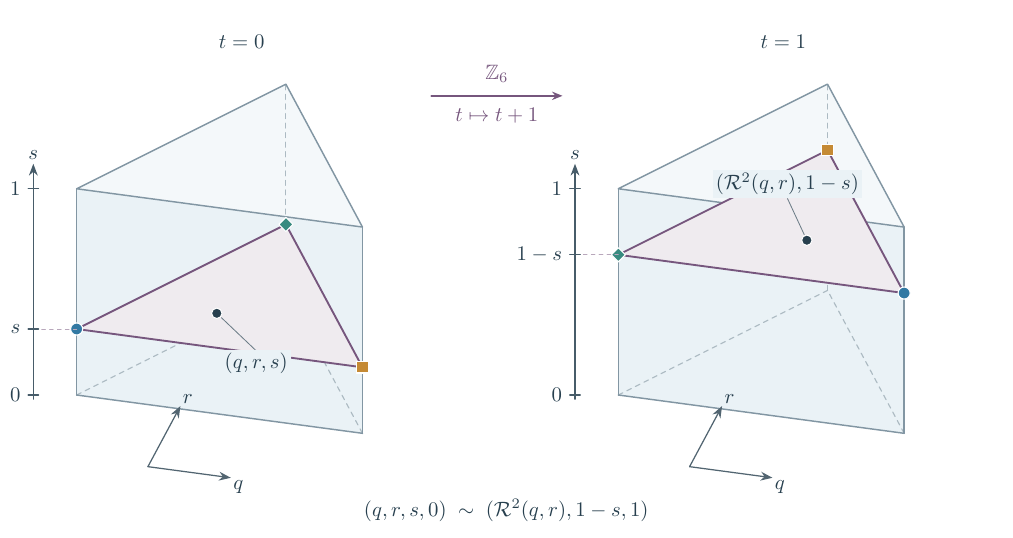}
    \caption{Boundary identification for the $\mathbb Z_6$ quotient. The triangular prisms represent the $t=0$ and $t=1$ sections of the holonomy domain, each containing the full color and weak Weyl chambers. The gluing $(q,r,s,0)\sim(\mathcal R^2(q,r),1-s,1)$ combines a cyclic color rotation with a weak twist. The black points denote
    a generic pair of identified points.}
    \label{fig:sm-z6-domain}
\end{figure*}

\noindent\textbf{The $\mathbb Z_6$ quotient.---} Combining the $\mathbb Z_2$ generator with the inverse $\mathbb Z_3$ generator gives $(\omega\mathbf{1}_3,-\mathbf{1}_2,e^{i\pi/3})$, which induces $(q,r,s,t)\sim(q+1,r,s+1,t+1)$. Returning the coordinates of both $SU(3)_c$ and $SU(2)_L$ to their Weyl chambers gives
\begin{equation}
    (q,r,s,t)\sim(\mathcal R^2(q,r),1-s,t+1).
    \label{eq:sm-z6-identification}
\end{equation}
Its square and cube reproduce the $\mathbb Z_3$ and $\mathbb Z_2$ identifications, respectively. We may therefore retain the full color and weak chambers with $0\leq t<1$, gluing the boundary faces as $(q,r,s,0)\sim(\mathcal R^2(q,r),1-s,1)$. The simultaneous color rotation and weak reflection are illustrated in Fig.~\ref{fig:sm-z6-domain}.

The fundamental domains of thermal holonomies derived here are fixed by the global form of the gauge group, independently of the detailed form of the potential. In the next section, we analyze the effective potential for the thermal holonomies and use these identifications to distinguish gauge-equivalent minima from physically inequivalent vacua.

\section{Standard Model Vacua at High Temperature}
\label{sec:SMhighT}

We consider the one-loop SM Weiss potential~\cite{Gross:1980br, Weiss:1980rj, Weiss:1981ev} in the massless high-temperature approximation, with the vanishing Higgs vacuum expectation value and zero chemical potential.  
Following~\cite{KorthalsAltes:1994be,Bronoff:1997an}, the holonomy-dependent part is $\pi^2T^4 V(q,r,s,t)$, where $V$ is the sum of the entries in Table~\ref{tab:sm-weiss} plus $15/8$, ensuring $V(0,0,0,0)=0$; see also~\cite{Meisinger:2001fi} for a derivation. Next we discuss the shape of the potential $V(q, r, s, t)$ for each $G_\Gamma$ in Eq.~\eqref{eq:SM_global}.

\begin{table}[t]
\centering
\small
\setlength{\tabcolsep}{4pt}
\renewcommand{\arraystretch}{1.2}
\begin{tabular}{@{}lc@{}}
\hline\hline
Field & Contribution to $V(q,r,s,t)$ \\
\hline
$G^a$
& $\frac43\left[v(r)+v(q+r/2)+v(q-r/2)\right]$
\\[3pt]
\hline
$W^i$
& $\frac43v(s)$
\\[3pt]
\hline
$Y$
& $0$
\\[3pt]
\hline
$H$
& $\frac23\sum_{\sigma=\pm1}v(\sigma s/2+t/2)$
\\[3pt]
\hline
$Q_L$
& $\begin{aligned}
-2\sum_{\sigma=\pm1}\bigl[
&v(q/3+r/2+\sigma s/2+t/6+1/2)\\
&+v(q/3-r/2+\sigma s/2+t/6+1/2)\\
&+v(-2q/3+\sigma s/2+t/6+1/2)\bigr]
\end{aligned}$
\\[3pt]
\hline
$u_R$
& $\begin{aligned}
-2\bigl[&v(q/3+r/2+2t/3+1/2)\\
        &+v(q/3-r/2+2t/3+1/2)\\
        &+v(-2q/3+2t/3+1/2)\bigr]
\end{aligned}$
\\[3pt]
\hline
$d_R$
& $\begin{aligned}
-2\bigl[&v(q/3+r/2-t/3+1/2)\\
        &+v(q/3-r/2-t/3+1/2)\\
        &+v(-2q/3-t/3+1/2)\bigr]
\end{aligned}$
\\[3pt]
\hline
$L_L$
& $-2\sum_{\sigma=\pm1}v(\sigma s/2-t/2+1/2)$
\\[3pt]
\hline
$e_R$
& $-2v(-t+1/2)$
\\
\hline\hline
\end{tabular}
\caption{Contributions of the SM fields to $V(q,r,s,t)$. The index $\sigma=\pm1$ labels the two weak eigenphases. We have already included the three generations of fermions. $v(x)$ is the Weiss periodic potential $v(x)=\{x\}^2(1-\{x\})^2$ with $\{x\}=x-\lfloor x\rfloor$, where $\lfloor x\rfloor$ is the floor function.}
\label{tab:sm-weiss}
\end{table}

\noindent\textbf{The trivial quotient.---} Minimizing $V(q,r,s,t)$ over the physical domain of $G_{\{1\}}$ yields six \emph{distinct} global minima related by the exact $\mathbb{Z}_6$ center symmetry:
\begin{equation}
\begin{aligned}
(q,r,s,t)\in\bigl\{&
(0,0,0,0),\;
(1,0,1,1),\;
(\tfrac12,1,0,2),\\
&
(0,0,1,3),\;
(1,0,0,4),\;
(\tfrac12,1,1,5)
\bigr\}.
\end{aligned}
\label{eq:global-minima}
\end{equation}
This is consistent with the conventional wisdom that the center symmetry is spontaneously broken at high temperature.~\footnote{Indeed, the six global minima can be represented as $(k,0,k,k)$ with $k=0,\ldots,5$. The identification due to large gauge transformation and Weyl group (see in Section~\ref{sec:holonomy}) implies that $(\frac{1}{2},1,0,2)\sim (2,0,2,2),\quad (0,0,1,3) \sim (3,0,3,3), \quad (1,0,0,4) \sim (4,0,4,4), \quad (\frac{1}{2},1,1,5) \sim(5,0,5,5)$, which are related to $(0,0,0,0)$ through the $\mathbb{Z}_6$ center transformation $(q,r,s,t)\mapsto(q+1,r,s+1,t+1)$.} 
The potential also has 24 \emph{distinct} metastable minima. For each global minimum, four metastable minima share the same $(q,r,s)$ and are obtained by shifting the hypercharge coordinate as
\begin{equation}
t\longmapsto t\pm\tfrac97,
\qquad
t\longmapsto t\pm1.74843\ldots,
\end{equation}
with $t$ understood modulo six. Their coordinates and potential values are listed together with the global minima in Table~\ref{tab:sm-minima}. (A careful analysis can confirm that the minima identified here are exhaustive.)

\begin{table}[t]
\centering
\small
\setlength{\tabcolsep}{3pt}
\renewcommand{\arraystretch}{1.2}
\begin{tabular}{@{}cccc@{}}
\hline\hline
$(q,r,s)$ & Global & \multicolumn{2}{c}{Metastable} \\
\cline{3-4}
& $V=0$ & $V\simeq0.9862$ & $V\simeq0.9432$ \\
\hline
$(0,0,0)$
& $0$ & $9/7,\ 33/7$
& $1.7484,\ 4.2516$ \\
$(1,0,1)$
& $1$ & $16/7,\ 40/7$
& $2.7484,\ 5.2516$ \\
$(\tfrac12,1,0)$
& $2$ & $5/7,\ 23/7$
& $0.2516,\ 3.7484$ \\
$(0,0,1)$
& $3$ & $12/7,\ 30/7$
& $1.2516,\ 4.7484$ \\
$(1,0,0)$
& $4$ & $19/7,\ 37/7$
& $2.2516,\ 5.7484$ \\
$(\tfrac12,1,1)$
& $5$ & $2/7,\ 26/7$
& $0.7484,\ 3.2516$ \\
\hline\hline
\end{tabular}
\caption{Hypercharge coordinates $t$ of the 6 global and 24 metastable minima in the physical holonomy domain of
$G_{\{1\}}$, at the indicated $(q,r,s)$. The columns are grouped by the potential value. Rational values are exact, while decimal values are rounded to four decimal places.}
\label{tab:sm-minima}
\end{table}

\noindent\textbf{The $\mathbb Z_2$ quotient.---} The identification Eq.~\eqref{eq:sm-z2-identification} reduces the $6$ branches of minima to $3$ branches. As a result, within the domain of $G_{\mathbb{Z}_2}$ as in Eq.~\eqref{eq:sm-z2-domain}, there remain 3 \emph{distinct} global minima and 12 \emph{distinct} metastable minima. The surviving global minima ${(0,0,0,0), (1,0,1,1), (\tfrac12,1,0,2)}$ are related by the residual $\mathbb Z_6/\mathbb Z_2\simeq\mathbb Z_3$ center symmetry, which is spontaneously broken. Indeed, taking the quotient corresponds to gauging the electric one-form symmetry, which reduces the genuine Wilson lines~\cite{Gaiotto:2014kfa}.

\noindent\textbf{The $\mathbb Z_3$ quotient.---} The identification Eq.~\eqref{eq:sm-z3-identification} reduces the $6$ branches of minima to $2$ branches, where only the ones within the domain of $G_{\mathbb{Z}_3}$ (i.e., Eq.~\eqref{eq:sm-z3-domain}) remain \emph{distinct}. The surviving global minima ${(0,0,0,0), (1,0,1,1)}$ are related by the residual $\mathbb Z_6/\mathbb Z_3\simeq\mathbb Z_2$ center symmetry, which is spontaneously broken.

\noindent\textbf{The $\mathbb Z_6$ quotient.---} All six center-related global minima are gauge equivalent, leaving the unique global minimum $(0,0,0,0)$, while four \emph{distinct} metastable minima remain. Indeed, the electric one-form symmetry becomes trivial when the $\mathbb Z_6$ quotient is taken, so there is no remaining charge that distinguishes the center-related branches.

\section{Additional Matter and the Vacuum Structure}
\label{sec:matter}

In this section, we examine the effects of new particles beyond the SM on the effective potential of thermal holonomies. In particular, we highlight the complementary aspects of new particles that are fully neutral under the $\mathbb{Z}_6$ center symmetry versus those that violate it.

Particles neutral under the $\mathbb Z_6$ center can substantially modify the potential~\cite {Bronoff:1997an}, particularly in the massless approximation, where their contributions are not subject to Boltzmann suppression. Meanwhile, the potential preserves exactly the $\mathbb Z_6$ center symmetry, and the six-fold vacuum degeneracy accordingly. Since all four possible $G_\Gamma$ are compatible with $\mathbb Z_6$ center-neutral new particles, the physical vacuum degeneracy only depends on gauging the electric one-form symmetry, as discussed in Section~\ref{sec:SMhighT}.

Particles charged under the $\mathbb{Z}_6$ center are exotic. Depending on the quantum numbers, they explicitly break the electric one-form symmetry of the SM (or a subgroup of it)~\cite{Li:2024nuo, Alonso:2024pmq, Koren:2024xof}, and their mass scale $M$ is likely to be much larger than the reheating temperature $T_{\rm RH}$ to avoid overproduction. This motivates us to consider the large-mass limit $T\leq T_{\rm RH}\ll M$~\cite{Meisinger:2001fi, Kashiwa:2012wa}, it renders the contribution to the periodic Weiss potential of order
\begin{equation}
\frac{(M/T)^{3/2}e^{-M/T}}{\sqrt2\,\pi^{7/2}}\sim 10^{-43} \quad\text{when}\quad M/T\sim 100.
\end{equation}
For example, the $(\mathbf3,\mathbf2)_Y$ representations with $Y=1/2,2/3,0$ discussed in~\cite{Li:2024nuo} preserve only $\mathbb Z_2,\mathbb Z_3,\{1\}$ subgroup of the
center, respectively; any gauged subgroup $\Gamma$ must lie within the surviving unbroken center.
For $M/T\gg1$, their exponentially suppressed contributions may not lift the original metastable minima, while global minima whose degeneracy is no longer protected can be lifted to metastable states. The number of physically distinct global minima also depends on the $\Gamma$ being gauged.

\section{Concluding remarks}
\label{sec:conclusions}

We have shown that gauging different subgroups of the SM electric one-form symmetry leaves the local thermal potential unchanged, but changes the physical holonomy domain and, consequently, the classification of its high-temperature vacua. This vacuum identification of apparently distinct vacua is analogous to the Lazarides-Shafi mechanism~\cite{Lazarides:1982tw} for resolving the axion domain-wall problem, recently clarified through a topological quantum field theory (TQFT)~\cite{Suzuki:2026xvf} (see also~\cite{Lu:2023ayc, Chatterjee:2019rch}). Since gauging the SM electric one-form symmetry can also be implemented by coupling to a TQFT~\cite{Kapustin:2014gua, Gaiotto:2014kfa}, it would be interesting to explore this connection more systematically. Furthermore, one may also connect the conventional Polyakov loop model to the Landau-Ginzburg description for spontaneously broken one-form symmetry~\cite{Iqbal:2021rkn}. 

Another interesting direction is to investigate the dynamical evolution of the possible string-wall networks. The four connected SM gauge groups $G_\Gamma$ (with or without additional matter) impose different gauge identifications among the vacua, which is only \emph{kinematics}. However, it may lead to distinct string-wall \emph{dynamics} and observational signatures for distinguishing $G_\Gamma$, even without discovering center-charged exotic particles. A detailed understanding of the defect formation and dynamics is warranted. 
One can also extend our analysis to the additional dark gauge groups, where the quotient center elements may have a nontrivial embedding between the visible and dark sectors, which was dubbed as topological mixing~\cite{Dierigl:2024cxm}.

Of course, we have only considered the high-temperature regime of the SM with unbroken electroweak symmetry, whereas toward the infrared there are both electroweak symmetry breaking~\cite{Sakamoto:2013gqa} and QCD confinement~\cite{Fukushima:2017csk}, which make the analysis less trivial.

\section{Acknowledgment}
I would first like to thank Motoo Suzuki for helpful discussions on the TQFT interpretation of vacuum identification in solving the axion domain-wall problem, Joan Elias Miro for suggesting references on the Landau-Ginzburg description for analyzing spontaneously broken one-form symmetries. 
We acknowledge the use of OpenAI's GPT-6 Astra for assistance with manuscript presentation and certain calculations. 
The work of L.X.X. is partially supported by European Research Council (ERC) grant n.101039756.

\bibliography{thermalSM.bib}

@article{Aharony:2013hda,
    author = "Aharony, Ofer and Seiberg, Nathan and Tachikawa, Yuji",
    title = "{Reading between the lines of four-dimensional gauge theories}",
    eprint = "1305.0318",
    archivePrefix = "arXiv",
    primaryClass = "hep-th",
    reportNumber = "WIS-03-13-APR-DPPA, WIS/03/13-APR-DPPA, UT-13-15, IPMU13-0081",
    doi = "10.1007/JHEP08(2013)115",
    journal = "JHEP",
    volume = "08",
    pages = "115",
    year = "2013"
}

@article{Aharony:2013kma,
    author = "Aharony, Ofer and Razamat, Shlomo S. and Seiberg, Nathan and Willett, Brian",
    title = "{3$d$ dualities from 4$d$ dualities for orthogonal groups}",
    eprint = "1307.0511",
    archivePrefix = "arXiv",
    primaryClass = "hep-th",
    reportNumber = "WIS-07-13-JUN-DPPA",
    doi = "10.1007/JHEP08(2013)099",
    journal = "JHEP",
    volume = "08",
    pages = "099",
    year = "2013"
}

@article{Tong:2017oea,
    author = "Tong, David",
    title = "{Line Operators in the Standard Model}",
    eprint = "1705.01853",
    archivePrefix = "arXiv",
    primaryClass = "hep-th",
    doi = "10.1007/JHEP07(2017)104",
    journal = "JHEP",
    volume = "07",
    pages = "104",
    year = "2017"
}

@article{Li:2024nuo,
    author = "Li, Hao-Lin and Xu, Ling-Xiao",
    title = "{Understanding the SM gauge group from SMEFT}",
    eprint = "2404.04229",
    archivePrefix = "arXiv",
    primaryClass = "hep-ph",
    doi = "10.1007/JHEP07(2024)199",
    journal = "JHEP",
    volume = "07",
    pages = "199",
    year = "2024"
}

@article{Davighi:2019rcd,
    author = "Davighi, Joe and Gripaios, Ben and Lohitsiri, Nakarin",
    title = "{Global anomalies in the Standard Model(s) and Beyond}",
    eprint = "1910.11277",
    archivePrefix = "arXiv",
    primaryClass = "hep-th",
    doi = "10.1007/JHEP07(2020)232",
    journal = "JHEP",
    volume = "07",
    pages = "232",
    year = "2020"
}

@article{Wan:2019gqr,
    author = "Wan, Zheyan and Wang, Juven",
    title = "{Beyond Standard Models and Grand Unifications: Anomalies, Topological Terms, and Dynamical Constraints via Cobordisms}",
    eprint = "1910.14668",
    archivePrefix = "arXiv",
    primaryClass = "hep-th",
    doi = "10.1007/JHEP07(2020)062",
    journal = "JHEP",
    volume = "07",
    pages = "062",
    year = "2020"
}

@article{Davighi:2020bvi,
    author = "Davighi, Joe and Lohitsiri, Nakarin",
    title = "{Anomaly interplay in $U(2)$ gauge theories}",
    eprint = "2001.07731",
    archivePrefix = "arXiv",
    primaryClass = "hep-th",
    doi = "10.1007/JHEP05(2020)098",
    journal = "JHEP",
    volume = "05",
    pages = "098",
    year = "2020"
}

@article{Anber:2021upc,
    author = "Anber, Mohamed M. and Poppitz, Erich",
    title = "{Nonperturbative effects in the Standard Model with gauged 1-form symmetry}",
    eprint = "2110.02981",
    archivePrefix = "arXiv",
    primaryClass = "hep-th",
    doi = "10.1007/JHEP12(2021)055",
    journal = "JHEP",
    volume = "12",
    pages = "055",
    year = "2021"
}

@article{Choi:2023pdp,
    author = "Choi, Yichul and Forslund, Matthew and Lam, Ho Tat and Shao, Shu-Heng",
    title = "{Quantization of Axion-Gauge Couplings and Noninvertible Higher Symmetries}",
    eprint = "2309.03937",
    archivePrefix = "arXiv",
    primaryClass = "hep-ph",
    reportNumber = "YITP-SB-2023-27, MIT-CTP/5606",
    doi = "10.1103/PhysRevLett.132.121601",
    journal = "Phys. Rev. Lett.",
    volume = "132",
    number = "12",
    pages = "121601",
    year = "2024"
}

@article{Reece:2023iqn,
    author = "Reece, Matthew",
    title = "{Axion-gauge coupling quantization with a twist}",
    eprint = "2309.03939",
    archivePrefix = "arXiv",
    primaryClass = "hep-ph",
    doi = "10.1007/JHEP10(2023)116",
    journal = "JHEP",
    volume = "10",
    pages = "116",
    year = "2023"
}

@article{Cordova:2023her,
    author = "Cordova, Clay and Hong, Sungwoo and Wang, Lian-Tao",
    title = "{Axion domain walls, small instantons, and non-invertible symmetry breaking}",
    eprint = "2309.05636",
    archivePrefix = "arXiv",
    primaryClass = "hep-ph",
    doi = "10.1007/JHEP05(2024)325",
    journal = "JHEP",
    volume = "05",
    pages = "325",
    year = "2024"
}

@article{vanBeest:2023mbs,
    author = "van Beest, Marieke and Boyle Smith, Philip and Delmastro, Diego and Mouland, Rishi and Tong, David",
    title = "{Fermion-monopole scattering in the Standard Model}",
    eprint = "2312.17746",
    archivePrefix = "arXiv",
    primaryClass = "hep-th",
    doi = "10.1007/JHEP08(2024)004",
    journal = "JHEP",
    volume = "08",
    pages = "004",
    year = "2024"
}

@article{Wang:2021ayd,
    author = "Wang, Juven and Wan, Zheyan and You, Yi-Zhuang",
    title = "{Cobordism and deformation class of the standard model}",
    eprint = "2112.14765",
    archivePrefix = "arXiv",
    primaryClass = "hep-th",
    doi = "10.1103/PhysRevD.106.L041701",
    journal = "Phys. Rev. D",
    volume = "106",
    number = "4",
    pages = "L041701",
    year = "2022"
}

@article{Wang:2020mra,
    author = "Wang, Juven",
    title = "{Ultra Unification}",
    eprint = "2012.15860",
    archivePrefix = "arXiv",
    primaryClass = "hep-th",
    doi = "10.1103/PhysRevD.103.105024",
    journal = "Phys. Rev. D",
    volume = "103",
    number = "10",
    pages = "105024",
    year = "2021"
}

@article{Alonso:2024pmq,
    author = "Alonso, Rodrigo and Dimakou, Despoina and West, Mia",
    title = "{Fractional-charge hadrons and leptons to tell the Standard Model group apart}",
    eprint = "2404.03438",
    archivePrefix = "arXiv",
    primaryClass = "hep-ph",
    reportNumber = "IPPP/24/13",
    doi = "10.1016/j.physletb.2025.139354",
    journal = "Phys. Lett. B",
    volume = "863",
    pages = "139354",
    year = "2025"
}

@article{Khoze:2024hlb,
    author = "Khoze, Valentin V.",
    title = "{Monopoles and fermions in the Standard Model}",
    eprint = "2405.18689",
    archivePrefix = "arXiv",
    primaryClass = "hep-ph",
    reportNumber = "IPPP/24/VVK, IPPP/24/78",
    doi = "10.1007/JHEP09(2024)146",
    journal = "JHEP",
    volume = "09",
    pages = "146",
    year = "2024"
}

@article{Koren:2024xof,
    author = "Koren, Seth and Martin, Adam",
    title = "{Fractionally charged particles at the energy frontier: The SM gauge group and one-form global symmetry}",
    eprint = "2406.17850",
    archivePrefix = "arXiv",
    primaryClass = "hep-ph",
    doi = "10.21468/SciPostPhys.18.1.004",
    journal = "SciPost Phys.",
    volume = "18",
    number = "1",
    pages = "004",
    year = "2025"
}

@article{Dierigl:2024cxm,
    author = "Dierigl, Markus and Novi{\v{c}}i{\'c}, Du{\v{s}}an",
    title = "{The axion is going dark}",
    eprint = "2409.02180",
    archivePrefix = "arXiv",
    primaryClass = "hep-th",
    reportNumber = "LMU-ASC 14/24",
    doi = "10.1007/JHEP12(2024)104",
    journal = "JHEP",
    volume = "12",
    pages = "104",
    year = "2024"
}

@article{Cao:2024lwg,
    author = "Cao, Qing-Hong and Ge, Shuailiang and Liu, Yandong and Wang, Jun-Chen",
    title = "{Berry phase in axion physics, SM global structure, and generalized symmetries}",
    eprint = "2411.04749",
    archivePrefix = "arXiv",
    primaryClass = "hep-ph",
    doi = "10.1016/j.physletb.2026.140218",
    journal = "Phys. Lett. B",
    volume = "874",
    pages = "140218",
    year = "2026"
}

@article{Alonso:2025rkk,
    author = "Alonso, Rodrigo and Dimakou, Despoina and Ha, Yunji and Khoze, Valentin V.",
    title = "{Charge quantisation, monopoles and emergent symmetry in the Standard Model and its embeddings}",
    eprint = "2507.01777",
    archivePrefix = "arXiv",
    primaryClass = "hep-ph",
    reportNumber = "IPPP/25/42",
    doi = "10.1007/JHEP12(2025)121",
    journal = "JHEP",
    volume = "12",
    pages = "121",
    year = "2025"
}

@article{Anber:2025gvb,
    author = "Anber, Mohamed M.",
    title = "{Gauging the Standard Model 1-form symmetry via gravitational instantons}",
    eprint = "2509.22788",
    archivePrefix = "arXiv",
    primaryClass = "hep-th",
    doi = "10.1007/JHEP02(2026)225",
    journal = "JHEP",
    volume = "02",
    pages = "225",
    year = "2026"
}

@article{Herrero-Brocal:2025rkx,
    author = "Herrero-Brocal, Antonio and Perez-Soler, Javier and Vicente, Avelino",
    title = "{Beyond SU(N): U(3){\texttimes}U(2) as the underlying symmetry of the strong and electroweak interactions}",
    eprint = "2512.14839",
    archivePrefix = "arXiv",
    primaryClass = "hep-ph",
    doi = "10.1103/7nvk-q5w3",
    journal = "Phys. Rev. D",
    volume = "114",
    number = "3",
    pages = "035009",
    year = "2026"
}

@article{Delgado:2026lwj,
    author = "Delgado, Antonio and Koren, Seth",
    title = "{Quark-Lepton Color-Flavor Unification}",
    eprint = "2605.30413",
    archivePrefix = "arXiv",
    primaryClass = "hep-ph",
    month = "5",
    year = "2026"
}

@article{Anber:2015wha,
    author = "Anber, Mohamed M. and Poppitz, Erich",
    title = "{On the global structure of deformed Yang-Mills theory and QCD(adj) on $ {\mathrm{\mathbb{R}}}^3\times {\mathbb{S}}^1 $}",
    eprint = "1508.00910",
    archivePrefix = "arXiv",
    primaryClass = "hep-th",
    doi = "10.1007/JHEP10(2015)051",
    journal = "JHEP",
    volume = "10",
    pages = "051",
    year = "2015"
}

@article{Poppitz:2021cxe,
    author = "Poppitz, Erich",
    title = "{Notes on Confinement on R3 {\texttimes} S1: From Yang{\textendash}Mills, Super-Yang{\textendash}Mills, and QCD (adj) to QCD(F)}",
    eprint = "2111.10423",
    archivePrefix = "arXiv",
    primaryClass = "hep-th",
    doi = "10.3390/sym14010180",
    journal = "Symmetry",
    volume = "14",
    number = "1",
    pages = "180",
    year = "2022"
}

@article{Gaiotto:2014kfa,
    author = "Gaiotto, Davide and Kapustin, Anton and Seiberg, Nathan and Willett, Brian",
    title = "{Generalized Global Symmetries}",
    eprint = "1412.5148",
    archivePrefix = "arXiv",
    primaryClass = "hep-th",
    doi = "10.1007/JHEP02(2015)172",
    journal = "JHEP",
    volume = "02",
    pages = "172",
    year = "2015"
}

@article{KorthalsAltes:1994be,
    author = "Korthals Altes, Chris P. and Lee, Ki-Myeong and Pisarski, Robert D.",
    title = "{Phase of the Wilson line at high temperature in the standard model}",
    eprint = "hep-ph/9406264",
    archivePrefix = "arXiv",
    reportNumber = "CU-TP-632",
    doi = "10.1103/PhysRevLett.73.1754",
    journal = "Phys. Rev. Lett.",
    volume = "73",
    pages = "1754--1757",
    year = "1994"
}

@inproceedings{Bronoff:1997an,
    author = "Bronoff, S. and Farakos, K. and Dvali, G. and Korthals Altes, C. P.",
    title = "{Wilson line in high temperature particle physics}",
    booktitle = "{2nd International Conference on Strong and Electroweak Matter}",
    eprint = "hep-th/9708043",
    archivePrefix = "arXiv",
    pages = "192--212",
    month = "5",
    year = "1997"
}

@article{Vecchi:2025qie,
    author = "Vecchi, Luca",
    title = "{When CP requires $\bar{\theta}=0$, not $\bar{\theta}=\pi$}",
    eprint = "2507.10680",
    archivePrefix = "arXiv",
    primaryClass = "hep-ph",
    month = "7",
    year = "2025"
}

@article{Gross:1980br,
    author = "Gross, David J. and Pisarski, Robert D. and Yaffe, Laurence G.",
    title = "{QCD and Instantons at Finite Temperature}",
    reportNumber = "PRINT-80-0538 (PRINCETON)",
    doi = "10.1103/RevModPhys.53.43",
    journal = "Rev. Mod. Phys.",
    volume = "53",
    pages = "43",
    year = "1981"
}

@article{Weiss:1980rj,
    author = "Weiss, Nathan",
    title = "{The Effective Potential for the Order Parameter of Gauge Theories at Finite Temperature}",
    reportNumber = "UBC-81",
    doi = "10.1103/PhysRevD.24.475",
    journal = "Phys. Rev. D",
    volume = "24",
    pages = "475",
    year = "1981"
}

@article{Weiss:1981ev,
    author = "Weiss, Nathan",
    title = "{The Wilson Line in Finite Temperature Gauge Theories}",
    reportNumber = "Print-81-0743 (BRITISH COLUMBIA)",
    doi = "10.1103/PhysRevD.25.2667",
    journal = "Phys. Rev. D",
    volume = "25",
    pages = "2667",
    year = "1982"
}

@article{Fukushima:2017csk,
    author = "Fukushima, Kenji and Skokov, Vladimir",
    title = "{Polyakov loop modeling for hot QCD}",
    eprint = "1705.00718",
    archivePrefix = "arXiv",
    primaryClass = "hep-ph",
    doi = "10.1016/j.ppnp.2017.05.002",
    journal = "Prog. Part. Nucl. Phys.",
    volume = "96",
    pages = "154--199",
    year = "2017"
}

@article{Sakamoto:2013gqa,
    author = "Sakamoto, Makoto and Takenaga, K.",
    title = "{Standard model with new order parameters at finite temperature}",
    eprint = "1312.7052",
    archivePrefix = "arXiv",
    primaryClass = "hep-th",
    reportNumber = "KOBE-TH-13-11",
    doi = "10.1103/PhysRevD.89.105032",
    journal = "Phys. Rev. D",
    volume = "89",
    number = "10",
    pages = "105032",
    year = "2014"
}

@article{Smilga:1993vb,
    author = "Smilga, Andrei V.",
    title = "{Are Z(N) bubbles really there?}",
    reportNumber = "BUTP-93-3, BUTP-93-03",
    doi = "10.1006/aphy.1994.1073",
    journal = "Annals Phys.",
    volume = "234",
    pages = "1--59",
    year = "1994"
}

@article{Smilga:1996cm,
    author = "Smilga, Andrei V.",
    title = "{Physics of thermal QCD}",
    eprint = "hep-ph/9612347",
    archivePrefix = "arXiv",
    reportNumber = "TPI-MINN-96-23, NUC-MINN-96-21-T",
    doi = "10.1016/S0370-1573(97)00014-8",
    journal = "Phys. Rept.",
    volume = "291",
    pages = "1--106",
    year = "1997"
}

@article{Cheluvaraja:1998yu,
    author = "Cheluvaraja, Srinath",
    title = "{Mean field analysis of the SO(3) lattice gauge theory at finite temperature}",
    eprint = "hep-lat/9809164",
    archivePrefix = "arXiv",
    reportNumber = "TIFR-TH-98-36",
    month = "9",
    year = "1998"
}

@article{Meisinger:2001fi,
    author = "Meisinger, Peter N. and Ogilvie, Michael C.",
    title = "{Complete high temperature expansions for one loop finite temperature effects}",
    eprint = "hep-ph/0108026",
    archivePrefix = "arXiv",
    doi = "10.1103/PhysRevD.65.056013",
    journal = "Phys. Rev. D",
    volume = "65",
    pages = "056013",
    year = "2002"
}

@article{Hamada:2025cwu,
    author = "Hamada, Yuta and Mukaida, Kyohei and Uchida, Fumio",
    title = "{Symmetries of hot SM, magnetic flux baryogenesis from helicity decay}",
    eprint = "2507.01576",
    archivePrefix = "arXiv",
    primaryClass = "hep-ph",
    reportNumber = "KEK-TH-2737, IPMU25-0035",
    doi = "10.1007/JHEP01(2026)040",
    journal = "JHEP",
    volume = "01",
    pages = "040",
    year = "2026"
}

@article{Hsin:2024lya,
    author = "Hsin, Po-Shen and Gomis, Jaume",
    title = "{Detecting Standard Model Gauge Group from Generalized Fractional Quantum Hall Effect}",
    eprint = "2411.18160",
    archivePrefix = "arXiv",
    primaryClass = "hep-th",
    month = "11",
    year = "2024"
}

@article{Wan:2024kaf,
    author = "Wan, Zheyan and Wang, Juven and You, Yi-Zhuang",
    title = "{Topological Responses of the Standard Model Gauge Group}",
    eprint = "2412.21196",
    archivePrefix = "arXiv",
    primaryClass = "hep-th",
    month = "12",
    year = "2024"
}

@article{Kashiwa:2012wa,
    author = "Kashiwa, Kouji and Pisarski, Robert D. and Skokov, Vladimir V.",
    title = "{Critical endpoint for deconfinement in matrix and other effective models}",
    eprint = "1205.0545",
    archivePrefix = "arXiv",
    primaryClass = "hep-ph",
    doi = "10.1103/PhysRevD.85.114029",
    journal = "Phys. Rev. D",
    volume = "85",
    pages = "114029",
    year = "2012"
}

@article{Lazarides:1982tw,
    author = "Lazarides, George and Shafi, Q.",
    title = "{Axion Models with No Domain Wall Problem}",
    reportNumber = "RU82-B-27",
    doi = "10.1016/0370-2693(82)90506-8",
    journal = "Phys. Lett. B",
    volume = "115",
    pages = "21--25",
    year = "1982"
}

@article{Suzuki:2026xvf,
    author = "Suzuki, Motoo and Yokokura, Ryo",
    title = "{Lazarides-Shafi axion models as Dijkgraaf-Witten theories}",
    eprint = "2602.12345",
    archivePrefix = "arXiv",
    primaryClass = "hep-th",
    doi = "10.1103/hl5s-73lj",
    journal = "Phys. Rev. D",
    volume = "113",
    number = "9",
    pages = "L091902",
    year = "2026"
}

@article{Kapustin:2014gua,
    author = "Kapustin, Anton and Seiberg, Nathan",
    title = "{Coupling a QFT to a TQFT and Duality}",
    eprint = "1401.0740",
    archivePrefix = "arXiv",
    primaryClass = "hep-th",
    doi = "10.1007/JHEP04(2014)001",
    journal = "JHEP",
    volume = "04",
    pages = "001",
    year = "2014"
}

@article{Iqbal:2021rkn,
    author = "Iqbal, Nabil and McGreevy, John",
    title = "{Mean string field theory: Landau-Ginzburg theory for 1-form symmetries}",
    eprint = "2106.12610",
    archivePrefix = "arXiv",
    primaryClass = "hep-th",
    doi = "10.21468/SciPostPhys.13.5.114",
    journal = "SciPost Phys.",
    volume = "13",
    pages = "114",
    year = "2022"
}

@article{Lu:2023ayc,
    author = "Lu, Qianshu and Reece, Matthew and Sun, Zhiquan",
    title = "{The quality/cosmology tension for a post-inflation QCD axion}",
    eprint = "2312.07650",
    archivePrefix = "arXiv",
    primaryClass = "hep-ph",
    reportNumber = "MIT-CTP 5644",
    doi = "10.1007/JHEP07(2024)227",
    journal = "JHEP",
    volume = "07",
    pages = "227",
    year = "2024"
}

@article{Chatterjee:2019rch,
    author = "Chatterjee, Chandrasekar and Higaki, Tetsutaro and Nitta, Muneto",
    title = "{Note on a solution to domain wall problem with the Lazarides-Shafi mechanism in axion dark matter models}",
    eprint = "1903.11753",
    archivePrefix = "arXiv",
    primaryClass = "hep-ph",
    doi = "10.1103/PhysRevD.101.075026",
    journal = "Phys. Rev. D",
    volume = "101",
    number = "7",
    pages = "075026",
    year = "2020"
}


\end{document}